\documentclass[hyper,letterpaper]{JHEP3} 

\usepackage{epsfig,multicol,bbm}

\newcommand{\reef}[1]{(\ref{#1})}
\DeclareMathSymbol{\IR}{\mathbin}{AMSb}{"52}

\title{A Holographic Superconductor in an External Magnetic Field}

\author{{Tameem Albash, Clifford V. Johnson}\\
	Department of Physics and Astronomy\\ University of Southern California\\ Los Angeles, CA 90089\\
	E-mail: \email{albash@usc.edu},\email{johnson1@usc.edu}, }

\preprint{\arXivid{0804.3466}}	

\abstract{We study a system of a complex charged scalar coupled to a Reissner--Nordstr\"om black hole in 3+1 dimensional anti--de Sitter spacetime, neglecting back--reaction.  With suitable boundary conditions, the cases of a neutral and purely electric black hole have been studied in various limits and were shown to yield key elements of superconductivity in the dual 2+1 dimensional field theory, forming a condensate below a critical temperature. By adding magnetic charge to the black hole, we immerse the superconductor into an external magnetic field. We show that a family of condensates can form and we examine their structure.  For finite magnetic field, they are localized in one dimension with a profile that is exactly solvable, since it maps to the quantum harmonic oscillator.  As the magnetic field increases, the condensate shrinks in size, which is reminiscent of the Meissner effect.}


\begin{document} 
\section{Introduction}
Since the early days of
AdS/CFT\cite{Maldacena:1998re,Gubser:1998bc,Witten:1998qj} (see ref.
\cite{Aharony:1999t} for a review) it has been tempting to consider
applying holographic duality to the study of important strongly
coupled phenomena in condensed matter systems, and superconductivty
has been high on the list\footnote{The fact that certain black holes
  and branes are known to exhibit a sort of Meissner
  effect\cite{Wald:1974np,Bicak:1980du,Chamblin:1998qm} at zero
  temperature has always added to the motivation, although it has not
  been clear how exactly this could be connected to a dual
  superconductivity.}. Since finite temperature in these duals
typically implies the presence of a black hole, and since
superconductivity requires a condensate to form below a certain
critical temperature, the existence of a holographically dual
background would seem to require a circumvention of various statements
of no--hair theorems (which go back to Wheeler\cite{Ruffini} --- for a
review, see ref.\cite{Bekenstein:1996pn}).  Generically, the black
holes would need to have some kind of scalar hair in order to be dual
to a superconductor (the scalar's asymptotic value would be the
condensate vacuum expectation value (vev)).

In a series of
studies\cite{Gubser:2005ih,Gubser:2008px,Gubser:2008zu}, Gubser has
presented a case for just the right kind of no--scalar--hair theorem evasion
to allow for a superconductor's dual to exist. The statement seems to
be that there do exist solutions that allow for a condensing scalar to
be coupled to the black hole if the charge on the black hole is large
enough. The scalar couples to (at least) a $U(1)$ under which the
black hole is charged, and its condensation breaks the gauge
symmetry spontaneously, giving a mass to the gauge field. In
particular, if the effective mass in the bulk of the scalar is
negative enough, the scalar field develops a non--trivial vev at the
boundary, giving the gauge field a non--zero mass.

Studying such solutions is hard to do, since the full equations are
 coupled and non--linear, and so numerical methods, and a number
of limits, have been employed in order to extract the key physics.
Gubser has studied\cite{Gubser:2008zu} the case of non--Abelian
Reissner--Nordstr\"om black holes condensing, and in a simpler model
that seems to capture some of the essentials in a limit, Hartnoll
et.al.,\cite{Hartnoll:2008vx} have studied a neutral black hole with a
charged scalar and Maxwell sector that do not back react on the
geometry. The latter authors have explored (with the aid of that
simplifying limit) some of the phenomenology of the condensate as a
function of temperature and shown that it maps rather well (where the
limit can be trusted) to familiar features of superconductivity in the
dual 2+1 dimensional theory.

Emboldened by these studies, we explored the case of adding an
external magnetic field to the system, to see how the condensate
behaves\footnote{As we were preparing this manuscript, a paper on the
  same subject (ref.\cite{Nakano:2008xc}) appeared on the
  Ar$\chi$iv.}. We have a fully back--reacted electrically and
magnetically charged Reissner--Nordstr\"om black hole, and a charged
scalar whose back--reaction we neglect in our computations. Since the
scalar does not back--react, we cannot hope to see all of the
signature physics of a superconductor in the presence of magnetism, as
the superconductor is not able to repel the background magnetic field.
Instead, we find that the condensate generically adjusts itself so as
to fill only a strip of finite width in the plane, thereby reducing
the total magnetic field that threads it. Remarkably, we can solve
exactly for the profile that it adopts, and we find that, as the magnetic field approaches infinity, the condensate shrinks to zero size.

\section{The Background}
We begin by introducing a charged, complex scalar field into the four
dimensional Einstein--Maxwell action with a negative cosmological
constant\footnote{We are using the mostly positive signature
  convention.}:
\begin{equation}\label{eqt:action}
S = \frac{1}{2 \kappa_4^2} \int d^4 x \sqrt{-G} \left\{ R + \frac{6}{L^2} + L^2 \left(-\frac{1}{4} F^2 - \left| \partial \Psi - i g A \Psi \right|^2 - V \left(\left|\Psi \right| \right)\right) \right\} \ .
\end{equation}
This action contains a term proportional to $A_\mu A^\mu \bar{\Psi}
\Psi$.  This term contributes negatively to the effective mass of the
charged scalar since the charged black hole will source $A_t$.  It is
exactly this term that allows (but does not guarantee) a non--trivial
vev for the scalar field to form.  When a vev is formed, by the usual
Higgs--Anderson mechanism, the gauge field develops a mass term
proportional to $A_\mu A^\mu \langle \bar{\Psi} \Psi \rangle$.
In the limit where the scalar field $\Psi$ does not back--react on the
geometry, the solution for the background geometry we take is that of
the dyonic black hole \cite{Romans:1991nq}:
\begin{eqnarray} \label{eqt:metric}
ds^2 &=& \frac{L^2 \alpha^2}{z^2} \left( - f \left(z \right) dt^2 + dx^2 + dy^2 \right) +\frac{L^2}{z^2} \frac{d z^2}{f \left(z \right)} \ , \nonumber \\
F &=& 2 h \alpha^2 dx \wedge dy + 2 q \alpha dz \wedge dt\ , \nonumber \\
f \left(z \right) &=& 1 + \left( h^2 + q^2 \right) z^4 - \left(1 + h^2 + q^2 \right) z^3 = \left(1-z \right)\left( z^2 +z+ 1 - \left(h^2 +q^2\right) z^3 \right) \ .
\end{eqnarray}
In the coordinate system used in equation \reef{eqt:metric}, $z$ is a
dimensionless radial coordinate scaled so that the event horizon of
the black hole is located at $z_h = 1 $ and the AdS boundary is at $ z
\to 0$.  The parameters $\alpha, h$, and $q$ are related to the mass,
magnetic charge, and electric charge of the black hole respectively,
but only $\alpha$ is dimensionful, with dimension of inverse length.
These quantities are in turn related to the temperature, external
magnetic field, and charge density of the charged adjoint matter in
the dual field theory.  The only other dimensionful parameter in the
solution is $L$, related to the AdS radius.  The Hawking temperature
 is given by the usual Gibbons--Hawking calculus\cite{Gibbons:1979xm}:
\begin{equation} \label{eqt:temp}
T = \frac{1}{\beta} =  \frac{\alpha}{4 \pi} \left( 3 - h^2 - q^2 \right) \ .
\end{equation}
Note that in order for the temperature to remain positive,
$\left(h^2+q^2\right) \leq 3$.  Saturating this inequality corresponds
to the extremal, zero--temperature case.  In order to determine the
effect of the magnetic and electric charges of the black hole, we
choose a particular form for the gauge field $A$, such that $ F = d
A$:
\begin{equation}
A = 2 h \alpha^2 x dy + 2 q \alpha \left( z - 1 \right) d t \ .
\end{equation}
We have explicitly added the pure gauge term $- 2 q \alpha d t$ in
order to have $A$ regular at the event horizon
\cite{Kobayashi:2006sb}.  The $A_y$ term provides a constant magnetic
field $B=F_{xy}$, and this is interpreted as corresponding to an
external magnetic field in the (2+1)--dimensional
system\cite{Hartnoll:2007ai}:
\begin{equation}
B = 2 h \alpha^2 \ .
\end{equation}
The $A_t$ term has two terms, a constant term and a term that goes to
zero at the boundary.  The constant term is interpreted as the
chemical potential (for an analogue of R--charge; see {\it e.g.},
refs\cite{Chamblin:1999tk,Chamblin:1999hg}), and the second term can
be related to the conjugate dual charge density of the theory {\it via}:
\begin{equation}
\rho =\frac{1}{\mathcal{V} \beta} \frac{\delta S_{\mathrm{on-shell}}}{\delta A_t (z = 0)} = - \frac{L^2}{\kappa_4^2} q \alpha^2 \ ,
\end{equation}
where $\mathcal{V}$ is the volume of the two--dimensional spatial
part of the field theory.  
\section{The Scalar Field}
\subsection{Review}
Let us review the results of ref.~\cite{Hartnoll:2008vx} to better clarify the relationship to our present work.  In that paper, the background is neutral, so both the electric and magnetic charge of the dyonic black hole have been set to zero.  Instead, the Maxwell--scalar sector is decoupled from the gravity sector by sending the coupling $g \to \infty$.  In order to see this, we must first rescale $A_\mu \to A_\mu /g$ and $\Psi \to \Psi/g$.  The Maxwell--scalar sector then has an overall $g^{-2}$, which when sent to infinity, decouples it from the gravity sector.  In this analysis, the potential is taken to be:
\begin{equation} \label{eqt:potential}
V \left(\left| \Psi \right| \right) = - 2 \bar{\Psi} \Psi / L^2 \ .
\end{equation}
Therefore, one can now study the Maxwell--scalar theory in the black hole background with Lagrangian:
\begin{equation}
\mathcal{L} = -\frac{1}{4} F^2 - \left| \partial \Psi - i  A \Psi \right|^2 + 2 \bar{\Psi} \Psi / L^2
\end{equation}
The equation of motion for the fields $\Psi$ and $A_\mu$ are:
\begin{eqnarray}
& \frac{1}{\sqrt{-G}} \partial_\mu \left( \sqrt{-G} G^{\mu \nu} \left( \partial_\nu \Psi - i  A_\nu \Psi \right) \right) + \frac{2}{L^2} \Psi - i  G^{\mu \nu} A_\mu \left(\partial_\nu \Psi - i A_\nu \Psi \right) = 0 \ ,& \label{eqt:complex} \\
& \frac{1}{\sqrt{-G}} \partial_\nu \left(\sqrt{-G} G^{\nu \lambda } G^{\mu \sigma} F_{\lambda \sigma} \right)  - G^{\mu \nu}  \left(i \left(\bar{\Psi} \partial_\nu \Psi - \partial_\nu \bar{\Psi} \Psi  \right) +2 A_\nu \bar{\Psi} \Psi \right)= 0 \ , &
\end{eqnarray}
and that of $\bar{\Psi}$ is simply the complex conjugate of equation \reef{eqt:complex}.  We take the ansatz:
\begin{equation} \label{eqt:ansatz}
 \Psi \equiv \Psi \left( z \right) = \tilde{\Psi}(z) /L \ , \quad A_t \equiv A_t \left(z \right) = \alpha \tilde{A}_t (z) \ ,
\end{equation}
where $\tilde{\Psi}$ and $\tilde{A}_t$ are dimensionless fields.  It is then consistent to take the phase of $\Psi$ to be constant.  All other fields are set to zero.  Under this ansatz, the equations of motion simplify to:
\begin{equation}
\partial_z^2 \tilde{\Psi}  + \left(\frac{f'}{f}-\frac{2}{z} \right) \partial_z \tilde{\Psi} + \frac{1}{  f^2  } \tilde{\Psi} \tilde{A}_t^2 + \frac{2}{z^2 f}\tilde{\Psi}= 0 \ , \quad \partial_z^2 \tilde{A}_t - \frac{2}{z^2 f} \tilde{\Psi}^2 \tilde{A}_t = 0 \ . 
\end{equation}
Without presenting the details of the analysis (see ref.~\cite{Hartnoll:2008vx}), we show in figure \ref{fig:O2_v1} the results of the variation of an order parameter as the temperature changes. The onset of superconductivity occurs for $T<T_c$. The critical temperature $T_c$ is proportional to the square root of the charge density.
\FIGURE[h]{\epsfig{file=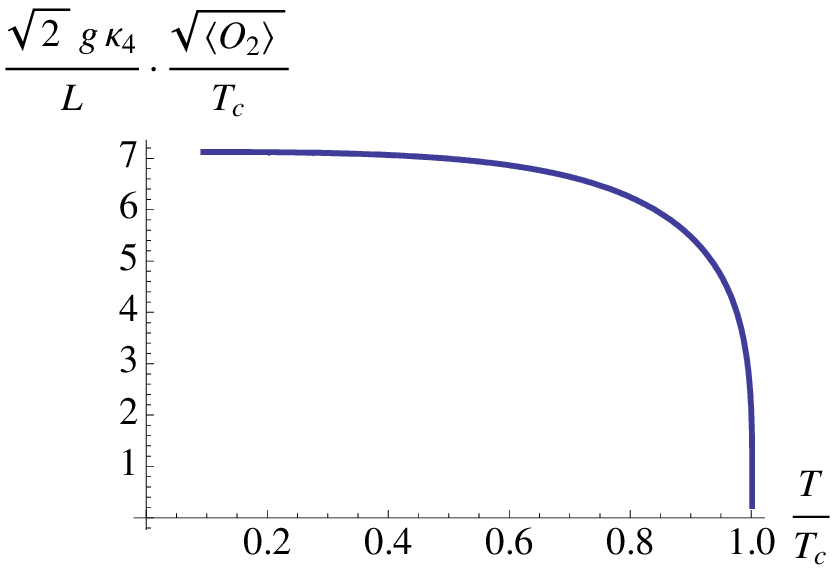,width=9cm}
 \caption{\small The vev of the $\Delta = 2$ operator as a function of the temperature for the isolated Maxwell--scalar sector studied in ref.~\cite{Hartnoll:2008vx}.  $T_c$ is proportional to the square root of the charge density.  Note that we use a different normalization, which accounts for the difference of a factor of $\sqrt{2}$ with ref.~\cite{Hartnoll:2008vx}.}
 \label{fig:O2_v1}}
\subsection{Perturbative limit}
We now consider the scalar field as a perturbation about the dyonic black hole background.  In this analysis, the Maxwell--scalar sector is not isolated from the gravity sector, since the Maxwell field has back--reacted on  the gravitational background.  We use the same potential considered in equation \reef{eqt:potential}, which corresponds to choosing $m^2 L^2 = -2 $ for the scalar field.  Before proceeding with our analysis, we would like to emphasize the relationship between this work and that of ref.~\cite{Hartnoll:2008vx}, which we reviewed in the previous section.  We work in the limit where the scalar does not backreact on the Maxwell fields, which should correspond approximately to taking $A_t \gg \Phi$ in ref.~\cite{Hartnoll:2008vx}.  From equation \reef{eqt:ansatz}, we see that this limit can be accomplished by taking $\alpha L \to \infty$.  In this limit, the charge density diverges, and hence the limit corresponds to taking $T/T_c \to 0$, \emph{i.e.} the left most end of the curve in figure \ref{fig:O2_v1}.  This argument is further established by the fact that, in this regime, $A_t$ in the coupled equations studied in ref.~\cite{Hartnoll:2008vx} and reviewed earlier behaves almost linearly.  In the dyonic black hole background, $A_t$ is linear.  This suggests that the analysis we propose captures the physics at $T/T_c \to 0$ in ref.~\cite{Hartnoll:2008vx}.  Therefore, for the physics we uncover, we are well below the critical temperature.  

The equation of motion for the scalar field is given by:
\begin{equation}\label{eqt:eom}
\frac{1}{\sqrt{-G}} \partial_\mu \left( \sqrt{-G} G^{\mu \nu} \left( \partial_\nu \Psi - i g A_\nu \Psi \right) \right) + \frac{2}{L^2} \Psi - i g G^{\mu \nu} A_\mu \left(\partial_\nu \Psi - i g A_\nu \Psi \right) = 0\ .
\end{equation}
The equation of motion for $\bar{\Psi}$ is simply the complex
conjugate of equation \reef{eqt:eom}. Using the fact that we only have
$A_t$ and $A_y$, and the only dependence is on the coordinates $x$ and
$z$, we consider an ansatz of the form $\Psi \equiv \Psi\left(x, z
\right)$.  The equation of motion simplifies to:
\begin{equation}
\frac{1}{\sqrt{-G}} \partial_z \left(\sqrt{-G} G^{zz} \partial_z \Psi \right) + G^{x x} \partial_x^2 \Psi + \frac{2}{L^2} \Psi - G^{yy} g^2 A_y^2\left(x \right) \Psi - G^{tt} g^2 A_t^2 \left(z \right) \Psi = 0\ .
\end{equation}
This equation and its complex conjugate are purely real.  Therefore,
the equations of motion imply that the phase of $\Psi$ is constant,
and so without loss of generality, we take $\Psi$ to be real.  We
assume a separable form for $\Psi$:
\begin{equation}
\Psi = X \left(x \right) Z \left(z \right) \ ,
\end{equation}
which further simplifies the equation of motion to:
\begin{eqnarray} \label{eqt:eom2}
&&\frac{1}{\sqrt{-G}} \partial_z \left(\sqrt{-G} G^{zz} Z' \left(z \right) \right) + \frac{2}{L^2} Z \left(z \right) - G^{tt} g^2 A_t^2 \left(z \right) Z \left(z \right)\\&&\hskip8.0cm + \frac{Z \left(z \right) G^{x x}}{X} \left( X''\left(x \right) - g^2 A_y^2\left(x \right) X\left(x\right)  \right)= 0 \ ,\nonumber
\end{eqnarray}
where we have used the fact that $G^{xx} = G^{yy}$.  In order for this
equation to be consistent, we must have that:
\begin{equation} \label{eqt:xeom}
 X''\left(x \right) -g^2 A_y^2\left(x \right) X\left(x\right) = - k^2 X \left(x \right) \ ,
\end{equation}
where $k^2$ is a constant.  By changing to a dimensionless variable $\tilde{x} = \sqrt{4 g h \alpha^2} x$ and setting $X \left(x \right) = \tilde{X} \left(\tilde{x} \right)$, equation \reef{eqt:xeom} can be brought to the form:
\begin{equation}
\tilde{X}'' \left(\tilde{x} \right) - \frac{\tilde{x}^2}{4} \tilde{X}\left(\tilde{x} \right) = - \frac{\tilde{k}^2}{2}  \tilde{X} \left(\tilde{x} \right) \ ,
\end{equation}
where $\tilde{k}^2 = k^2 / 2 g h \alpha^2$.  Generically, the solutions to this equation can be written in terms of confluent hypergeometric functions, but for the case when $\tilde{k}^2$ is an odd integer, the solutions can be written in terms of Hermite functions  $H_n$ of order $n=(\tilde{k}^2-1)/2$.  We restrict ourselves to this case since the Hermite functions decay exponentially for large~$x$, which seems to be the natural physical choice.  Henceforth in this paper, $\tilde{k}^2$ is taken to be an odd integer, and we display some examples of the functions in figure~\ref{fig:hermite}.  However, in our present setup, we have assumed that the phase of the scalar field remains constant.  Since only $\tilde{k}^2 = 1$ corresponds to a configuration that preserves its sign, it is the only physical solution for our ansatz.
\FIGURE{\epsfig{file=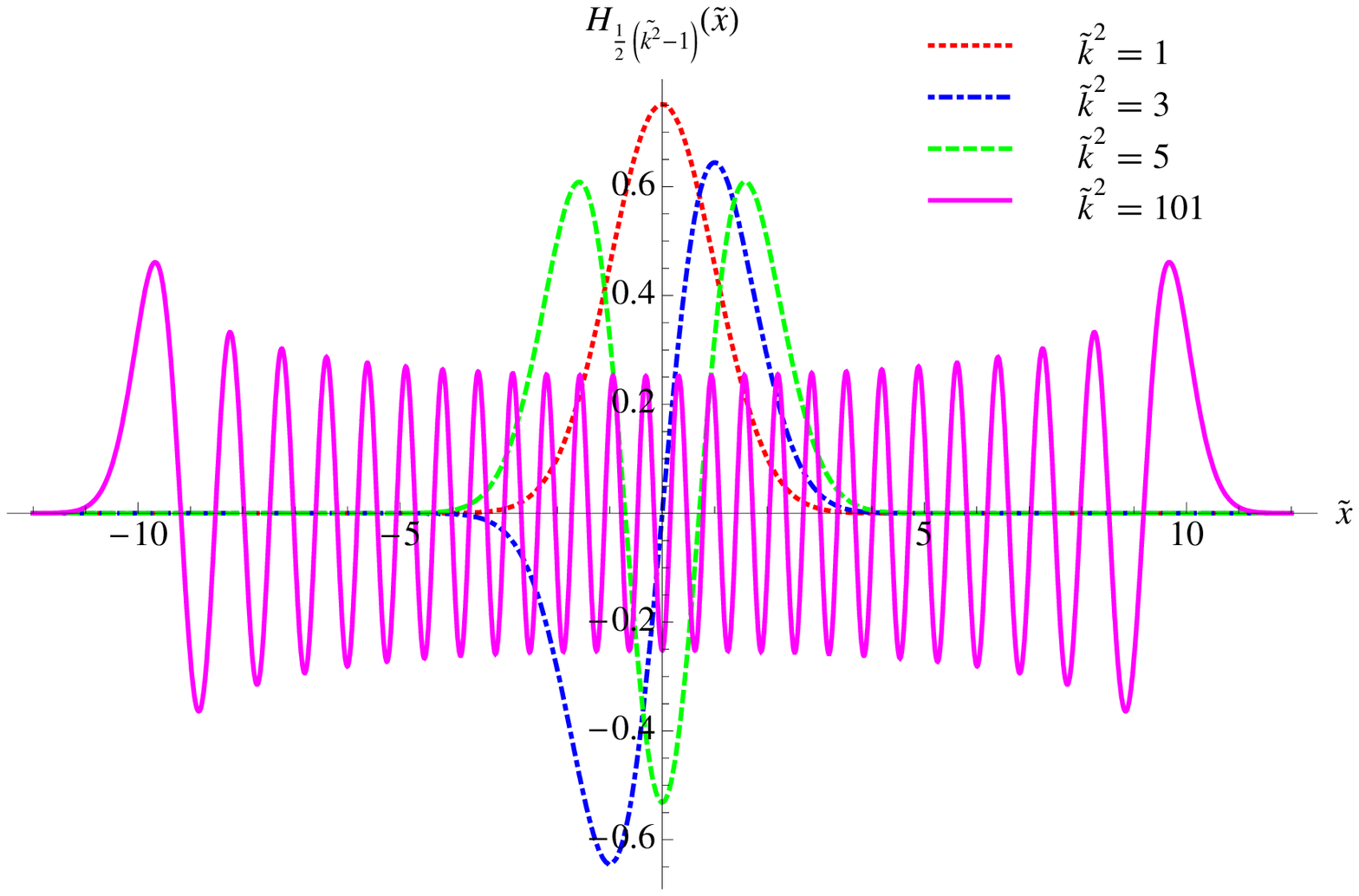,width=10cm}
 \caption{\small Some Hermite functions $H_n$ of order $n=(\tilde{k}^2-1)/2$, for $n=1,3,5$ and $101$.}
 \label{fig:hermite}}
As a result, our equation is exactly  Schr\"odinger's equation
for a simple harmonic oscillator!  Interestingly, the system confines
the $x$--extent of the condensate in a potential that is exactly
quadratic.  With the constant $B$--field passing through the $(x,y)$
plane, we can compute the flux of $B$ through a region as:
\begin{equation}
{\Phi} = B \Delta x \Delta y\ ,
\end{equation}
Using the dimensionless variables we introduced earlier, we have:
\begin{equation}
{\Phi} = 2h\alpha^2 \frac{\Delta \tilde{x}\Delta \tilde{y}}{4 \alpha^2 g h} = \frac{\Delta \tilde{x}\Delta \tilde{y}}{2 {g}} \ .
\end{equation}
$\Delta {\tilde x}$ only depends on $\tilde{k}^2$ since it is entirely
determined by the behavior of the Hermite function.  To estimate
$\Delta {\tilde x}$, we use that:
\begin{equation}
\langle \tilde{x}^2 \rangle = \int_{-\infty}^{\infty} d \tilde{x} \left|H_{n} \left(\tilde{x} \right)\right|^2 {\tilde x}^2 = n+ \frac{1}{2} = \frac{\tilde{k}^2}{2} \ , 
\end{equation}
and $\langle x \rangle=0$, then our flux becomes:
\begin{equation}
\frac{\Phi}{\ell} = \frac{1}{2g}  \sqrt{\frac{\tilde{k}^2}{2 }} \ .
\end{equation}
where ${\ell}$ is the extent in the $y$--direction, and since it is
infinite there, we consider the flux per unit $y$--length.

Substituting our result of equation~(\ref{eqt:xeom}) for the
$x$--dependence back into equation \reef{eqt:eom2}, the equation of
motion for $Z\left(z \right)$ becomes:
\begin{equation}
\frac{1}{\sqrt{-G}} \partial_z \left(\sqrt{-G} G^{zz} Z' \left(z \right) \right) + \frac{2}{L^2} Z \left(z \right) - G^{tt} g^2 A_t^2 \left(z \right) Z \left(z \right) - 2 g h \alpha^2 \tilde{k}^2 Z \left(z \right) G^{x x}= 0 \ ,
\end{equation}
Substituting in the functions, we get:
\begin{equation} \label{eqt:eomZ}
Z'' \left(z \right) + \left( \frac{f'\left(z \right) }{f\left(z \right) } - \frac{2}{z} \right) Z'\left(z \right) + \left(\frac{4 g^2 q^2}{f\left(z \right) ^2} \left(z-1 \right)^2 + \frac{2}{z^2 f\left(z \right) } - \frac{2 g h \tilde{k}^2}{f \left(z \right) } \right) Z\left(z \right) = 0 \ .
\end{equation}
It is interesting to note that all $\alpha$ and $L$ dependence in the equation have cancelled.  This would seem to indicate that the scalar field's behavior does not depend on $\alpha$ and $L$, but it is
important to remember that we are working in the perturbative limit where $\Psi$ is supposed to be small.  Earlier, we stated that to ensure the perturbative limit is consistent, we must have the quantity $\alpha L$ to be large.  Since the only dimensionful parameters are $\alpha$ and $L$, we take:
\begin{equation}
\Psi = \frac{1}{\alpha L^2} \tilde{\Psi} = \frac{1}{\alpha L^2} X\left(x \right) \tilde{Z} \left(z \right)\ ,
\end{equation}
which ensures that the scalar field is always perturbative, and we can now work in terms of the dimensionless function $\tilde{Z}$.  \\
Near the AdS boundary, equation \reef{eqt:eomZ} becomes:
\begin{equation} \label{eqt:boundary}
\tilde{Z}'' \left( z \right) - \frac{2}{z} \tilde{Z}' \left( z \right) + \frac{2}{z^2} \tilde{Z}\left( z \right) = 0 \ ,
\end{equation}
which has the  solution:
\begin{equation}
\lim_{z \to 0} \tilde{Z} \left( z \right) = \Psi_1 z + \Psi_2 z^2  \ ,
\end{equation}
where $\Psi_1$ and $\Psi_2$ are dimensionless constants.  Both of
these solutions are normalizable, so one is not the source of the
other.  Having chosen $m^2 L^2 = -2$ by our choice of potential (see
equation \reef{eqt:potential}), there is not a unique boundary
condition at the AdS boundary \cite{Klebanov:1999tb}.  $\Psi_1$ is
proportional to the vev of the $\Delta = 1$ operator ($\langle
\mathcal{O}_1 \rangle$) and $\Psi_2$ is proportional to the vev of the
$\Delta = 2$ operator ($\langle \mathcal{O}_2\rangle $) in the
boundary field theory, and only one of these is turned on by the
boundary condition.  This forces us to pick as one of our boundary
conditions:
\begin{equation} \label{eqt:bc}
\Psi_1 = 0 \ \mathrm{or} \  \Psi_2 = 0 \ .
\end{equation}
We choose to only work with the $\Delta = 2$, since we find the same
general qualitative behavior for both operators.  The exact
relationship between the vev and the asymptotic value of $\tilde{Z}$
can be calculated using the holographic dictionary (details are shown
in Appendix \ref{appendix:vev}):
\begin{equation}
\langle \mathcal{O}_2 \rangle =\frac{ \delta S_{\mathrm{on-shell}}}{\delta \Psi(z=0)} = \frac{L^2}{2 \kappa_4^2} \alpha^2  \Psi_2 \ .
\end{equation}
We can also study the behavior of the solution near the event horizon.
We find that there are three distinct possible cases.  First, if $ g h
\tilde{k}^2 < 1$, then the equation of motion becomes:
\begin{equation}
\tilde{Z}'' \left( z \right) - \frac{1}{1-z} \tilde{Z}'\left( z \right) + \frac{a^2}{1-z}\tilde{Z} \left( z \right) = 0 \ ,
\end{equation}
where
\begin{displaymath}
a^2 =  2 \frac{1 - 1 g h \tilde{k}^2 }{3 - h^2 - q^2} > 0 \ ,
\end{displaymath}
which has Bessel functions as solutions:
\begin{eqnarray}
\lim_{z \to 1} \tilde{Z} \left(z \right) &=& \psi_1 J_0 \left(2 a \sqrt{ \left(1-z\right)} \right) + \psi_2 Y_0 \left( 2 a \sqrt{\left(1-z\right)}  \right)  \nonumber \\ 
&\approx& \psi_1 +  \psi_2 \left(\frac{2}{\pi} \gamma + \frac{1}{\pi}  \log \left(a \left(1- z \right)\right) \right) \ ,
\end{eqnarray}
where $\gamma$ is the Euler--Mascheroni constant. Since we want the
field to be finite at the event horizon, we choose our other boundary
condition to be $\psi_2 = 0$.  For $ g h \tilde{k}^2 = 1$, the
equation of motion becomes
\begin{equation}
\tilde{Z}'' \left( z \right) - \frac{1}{1-z} \tilde{Z}'\left( z \right) +b^2 \tilde{Z} \left( z \right) = 0 \ ,
\end{equation}
where 
\begin{displaymath}
b^2 = \frac{4 g^2 q^2}{\left(3-h^2-q^2 \right)^2} + \frac{4}{3- h^2 -q^2} > 0 \ ,
\end{displaymath}
which has as solutions:
\begin{eqnarray}
\lim_{z \to 1} \tilde{Z} \left(z \right) &=& \psi_1 J_0 \left( b^2  \left(1-z\right) \right) + \psi_2 Y_0 \left( b^2 \left(1-z\right)  \right)  \nonumber \\ 
&\approx& \psi_1 +  \psi_2 \left(\frac{2}{\pi} \gamma + \frac{2}{\pi}  \log \left(b^2 \left(1- z \right)\right) \right) \ .
\end{eqnarray}
Since we want the field to be finite at the event horizon, we would
choose the same boundary condition as before, {\it i.e.} $\psi_2 =
0$.  Finally, if $ g h \tilde{k}^2 >1$, we have:
\begin{equation}
\tilde{Z}'' \left( z \right) - \frac{1}{1-z} \tilde{Z}'\left( z q\right) - \frac{a^2}{1-z}\tilde{Z} \left( z \right) = 0 \ ,
\end{equation}
where $a^2 = 2 \frac{1 - 1 g h \tilde{k}^2 }{3 - h^2 - q^2} > 0$,
which has as solutions:
\begin{eqnarray}
\lim_{z \to 1} \tilde{Z} \left(z \right) &=& \psi_1 I_0 \left(2 a \sqrt{ \left(1-z\right)} \right) + \psi_2 K_0 \left( 2 a \sqrt{\left(1-zq\right)}  \right)  \nonumber \\ 
&\approx& \psi_1 +  \psi_2 \left(- \gamma -  \log \left(a \left(1- z \right)\right) \right) \ .
\end{eqnarray}
Again, if we want the field to be finite at the event horizon, we would choose the same boundary condition as before, {\it i.e.}, $\psi_2 = 0$.
%
\section{Numerical Method and Results}
There are several parameters that we can vary in this problem, $g, \ 
\tilde{k}^2, \ q, \ h$.  For the following analysis, we take $g=1$ for
simplicity, but we can expect similar behavior for other values of
$g$.  In order to solve equation \reef{eqt:eomZ}, we use a shooting
method.  We impose the following initial conditions at the event
horizon:
\begin{equation} \label{eqt:IC}
\tilde{Z}(1) = 1 \ , \quad \tilde{Z}'(1) =  \frac{2 -2 g h \tilde{k}^2}{3- h^2 - q^2 }  \ .
\end{equation}
We find that these initial conditions do not necessarily satisfy
equation \reef{eqt:bc}.  For a given~$\tilde{k}^2$, only for certain
values of $h$ and $q$ do we get to satisfy the appropriate boundary
conditions.  Therefore, we fix the value of $h$, and then scan through
possible values of $q$ until the appropriate boundary condition at
$z=0$ is satisfied, exhibiting our condensate.  

We show in figure \ref{fig:psi} two allowed solutions for the scalar
field for a fixed $\tilde{k}^2$ and $h$ but with different $q$ values.
Ref.~\cite{Gubser:2008px} argues that only the zero--node solution
matters to the phase structure, so we consider only these solutions in
our subsequent analysis\footnote{Indeed, we find that it is only for
  these zero--node solutions that persist for low enough
  temperature.}.  It is important to emphasize that changing the value of the
scalar at the event horizon from the value given in equation
\reef{eqt:IC} does not change the required value of $h$ and $q$ for
the scalar to condense, but it does change the value of the vev.
Therefore, at  particular values of $h$ and $q$ that allow for a
condensate to form, there is a whole range of allowed vev values for
the operator depending on its value at the event horizon.
\FIGURE[h]{\epsfig{file=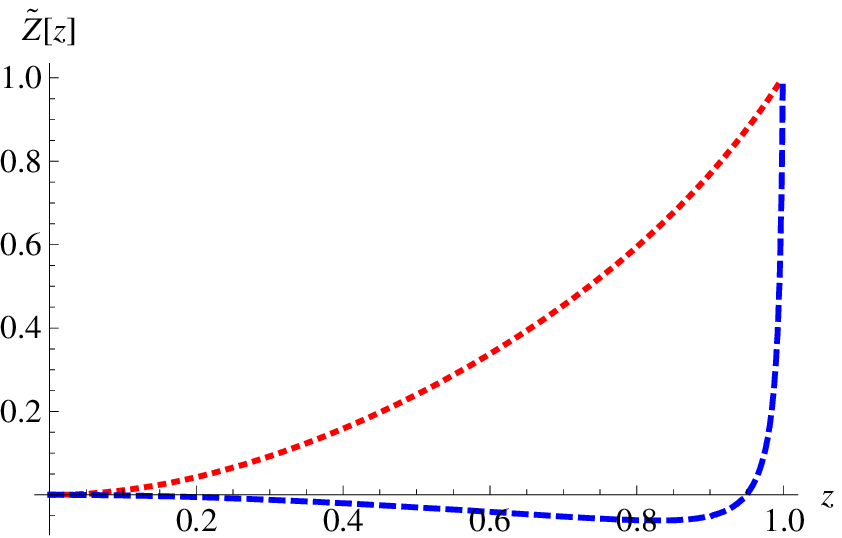,width=9cm}
 \caption{\small Two possible solutions for the scalar field with $\mathcal{O}_2$ turned on.  The parameters used are $\tilde{k}^2 = 1$ and $h=0.1$.}
 \label{fig:psi}}
For convenience, we define a temperature $\tilde{T}$ (with its
corresponding $\tilde{q}$ for which the scalar condenses) to be the
temperature at zero magnetic field ($h=0$) at which we find a solution.  This
is given by (\emph{cf} equation \reef{eqt:temp}):
\begin{equation}
\tilde{T} = \alpha \frac{3 - \tilde{q}^2}{4 \pi} \ .
\end{equation}
Using this definition, we can define several dimensionless quantities of interest:
\begin{equation} \label{eqt:results}
 \frac{\kappa_4}{L} \frac{\sqrt{\langle \mathcal{O}_2 \rangle}}{\tilde{T}} = \frac{ 4 \pi }{3-\tilde{q}^2} \sqrt{\frac{\Psi_2}{2}} \ ,  \quad \frac{B}{\tilde{T}^2} = \left(\frac{ 4 \pi}{3 - \tilde{q}^2}\right)^2 2 h , \quad - \frac{\kappa_4^2}{L^2} \frac{\rho}{\tilde{T}^2} =  \left(\frac{ 4 \pi}{3 - \tilde{q}^2}\right)^2 q \ .
\end{equation}
We present the results of our numerical condensate search in terms of
these quantities in figure~\ref{fig:results1}. 
\FIGURE{\epsfig{file=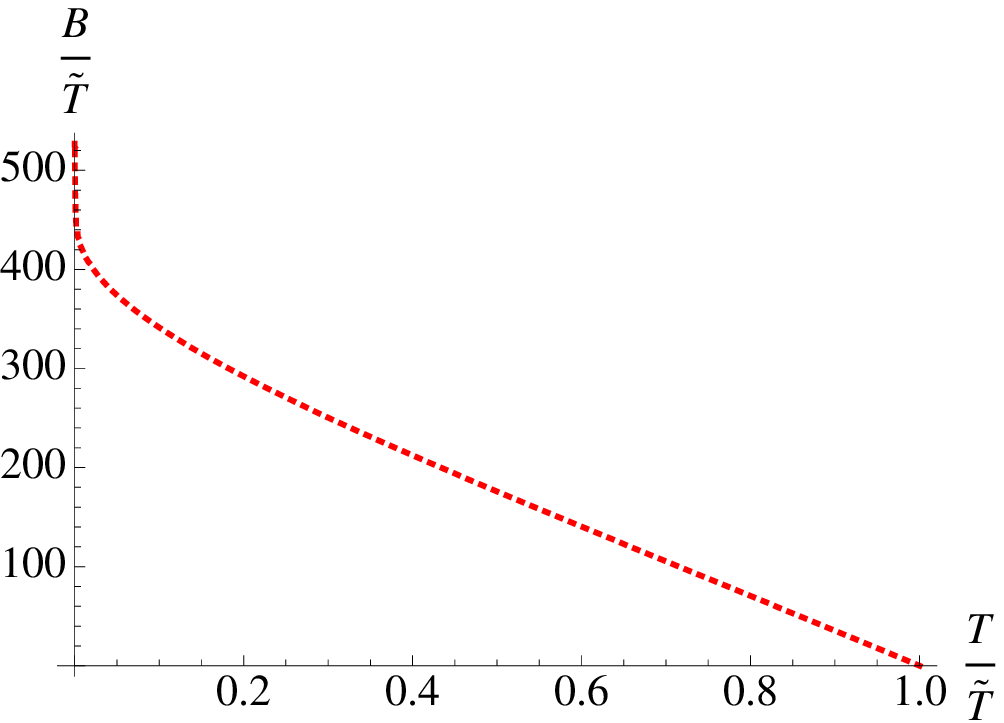,width=7.5cm} \epsfig{file=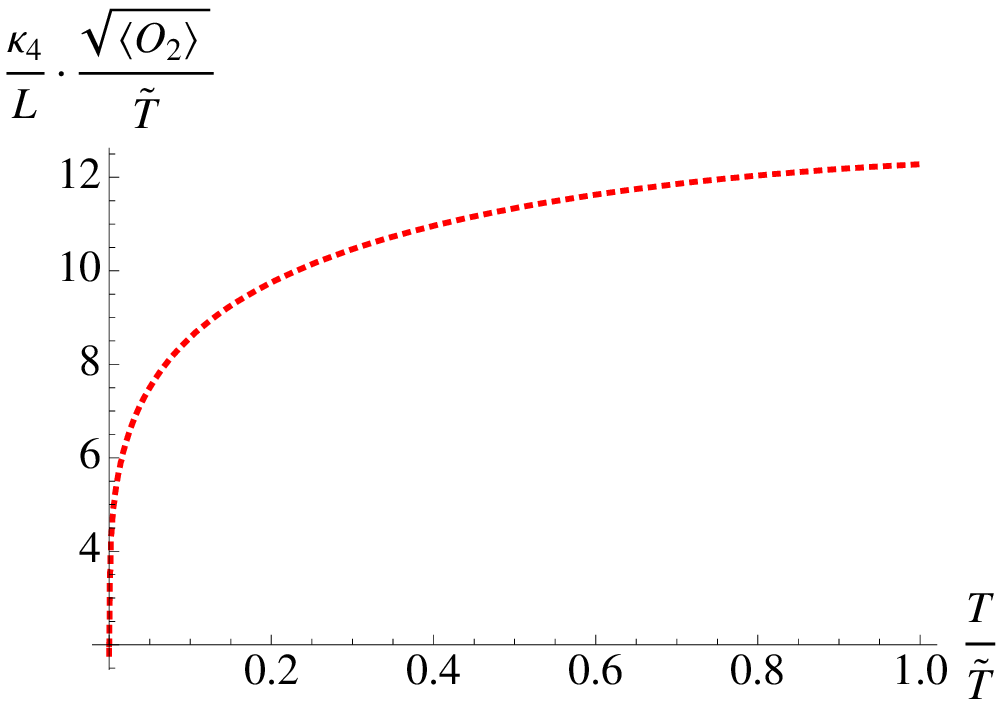,width=7.5cm}\epsfig{file=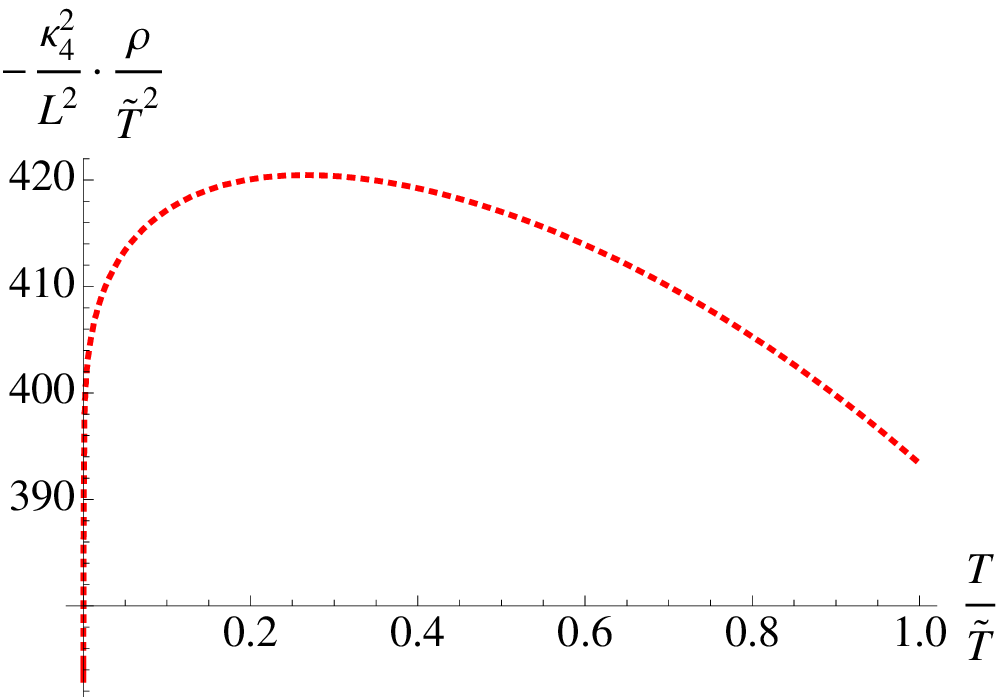,width=7.5cm}
 \caption{\small The graphs depict the following: the allowed $B$ and $T$ values for $\mathcal{O}_2$ to condense, the vev for $\mathcal{O}_2$, and the allowed $\rho$ and $T$ values for $\mathcal{O}_2$ to condense.}
 \label{fig:results1}}
%
\section{Discussion}

As previously alluded to, since we are working in a limit where the scalar is not back--reacting on the Maxwell field, we cannot (as in ref.\cite{Hartnoll:2008vx}) track the dependence of the vev on temperature all the way to the transition temperature\footnote{Note that ${\tilde T}$ is {\sl not} the transition temperature, but merely a normalization set by the solution at $h=0$.} $T_c$. We instead simply exhibit the values of the ratios of temperature, charge density,  and magnetic field to $\tilde{T}$ that allow a condensate to form in our study. This is enough to allow us to study the condensate's spatial behavior, and we are firmly below $T_c$ in all that we do..

There are several interesting features, as displayed in figure~\ref{fig:results1}.  Note that in ref.~\cite{Hartnoll:2008vx}, the scalar condenses at a particular value of $T/T_c$, with $T_c \propto \rho^{1/2}$.  Similarly, the scalar in our setup condenses at a particular value of $\rho/\tilde{T}^2$ and $B/\tilde{T}^2$.  There exists a minimum value of $\rho/\tilde{T}^2$ for the scalar to condense.  Beyond these values, condensation occurs but is not believed to be stable \cite{Gubser:2008px}, and the only allowed solution is the trivial one (see the previous section for more details).  

At zero magnetic field, the condensate fills the plane and requires the lowest ratio of $\tilde{\rho}/\tilde{T}^2$.  To see that it fills the plane, we recall that the profile in the $x$--direction for $\tilde{k}^2 = 1$ is given by:
\begin{equation}
X(x) = e^{-\frac{\tilde{x}^2}{4} } = e^{- g h \alpha^2 x^2} \ .
\end{equation}
Therefore, in the limit of $h \to 0$, we have $X(x) \to 1$, \emph{i.e.} there is no $x$--dependence.  As the magnetic field is turned on and increases (reading the first and the other two plots from right to left as $T/\tilde{T}$ decreases), the value of $\rho/\tilde{T}^2$ required for the scalar to condense steadily increases and the condensate has a finite thickness along the $x$--direction.  At around $T/\tilde{T} = 0.25$, the ratio $\rho/\tilde{T}^2$ needed drops rapidly.  The condensate also drops rapidly around the same interval, which suggests that the magnetic field might be overcoming the forces keeping the superconductor together.  However, we find that even at $T/\tilde{T} = 0$, the condensate persists with a standard deviation of $1/ \sqrt{6 g \alpha^2}$.

This result is of particular interest because, in the limit of large magnetic field, \emph{i.e.} $\alpha \to \infty$, the standard deviation approaches zero.  Therefore, the magnetic field, as it grows, shrinks the condensate away completely.  This is reminiscent of the Meissner effect, where the magnetic field expels the condensate.  The condensate itself cannot expel the magnetic field, as is usually the case, since the scalar cannot back--react on the background magnetic field.

It is interesting to speculate on what higher $\tilde{k}^2$ values could mean if for example the ansatz for the scalar field is modified to make them physical.  As an approximation, we can naively proceed with our setup with higher values of $\tilde{k}^2$.  We find that as ${\tilde k}^2$ increases and the $x$--width of the condensate expands (see figure~\ref{fig:hermite}), the corresponding $B$--field associated with it is smaller.  It is interesting to follow this to large ${\tilde k}^2$, the ``classical'' limit of the quantum harmonic oscillator controlling the $x$--profile: The ${\tilde k }^2\to\infty$ limit has the condensate filling the plane while the magnetic field $B\to0$. Pleasingly, this is consistent with the limit of small and large magnetic field we discussed above\footnote{In making sense of the $B$--field within the condensed region, where we expect it would be forced to zero in the fully back--reacting system, it is tempting to interpret the $x$ profile of the scalar quite literally and take $B$ times the average value within the region. This would mostly then integrate to zero in the interior of the sample, leaving only some non--zero contribution at the edges when $n=({\tilde k}^2-1)/2$ is even, and canceling exactly when $n$ is odd, but this is possibly too naive.}.

\section*{Acknowledgments}
TA thanks L'Institut de Physique Th\'eorique (IPhT), Saclay, France
for hospitality. He would also like to thank Nikolay Bobev and Hubert
Saleur for discussions.  This research was supported by the U.S.
Department of Energy.
\appendix
\section{Holographic Dictionary} \label{appendix:vev}
The solution of the equation of motion admits two normalizable
solutions at the AdS boundary, which we reproduce here:
\begin{equation} 
\lim_{z \to 0} \tilde{Z} \left( z \right) = \Psi_1 z + \Psi_2 z^2  \ , \nonumber
\end{equation}
To proceed with calculating the vev of the $\Delta=2$ operator
$\mathcal{O}_2$ using the holographic dictionary, the procedure is to
assume that $\Psi_1$ is the source of the operator
\cite{Klebanov:1999tb}.  To proceed, we first write the asymptotic
solution of the full scalar field as:
\begin{equation}
\lim_{z\to 0} \Psi(x,z) = e^{i \varphi} \left(\phi_0(x) z +  A(x) z^2\right) \ ,
\end{equation}
where we have explicitly shown the constant phase of the scalar field.
Next, we calculate the variation of the on--shell action:
\begin{eqnarray}
\delta S_{\mathrm{on-shell}} &=& - \frac{L^2}{2 \kappa_4^2} \int d^3 x \sqrt{-G} G^{zz} \partial_z \Psi(x,z) \delta \bar{\Psi}(x,z) \bigg|_{z=0}^{z=1} \nonumber \\
& = &\lim_{z \to 0} \frac{L^2}{2 \kappa_4^2} \int d^3 x \frac{L^2 \alpha^3}{z^2} \left(z \phi_0(x) \delta \phi_0(x) + 2 z^2 A(x) \delta \phi_0(x) + z^2 \phi_0(x) \delta A(x) + O(z^3) \right) \nonumber
\end{eqnarray}
At this point, we may worry about the divergence produced by the first
term in this expression, but we have not included the following
counterterm in the expression of our action:
\begin{equation}
S_{\mathrm{counter}} = - \frac{L^2}{4 \kappa_4^2} \sqrt{- \gamma} \frac{1}{L} \int d^3 x \bar{\Psi} (x,z = 0)\Psi (x,z = 0) \ ,
\end{equation}
where $\gamma$ is not the Euler--Mascheroni constant. Varying this
counterterm and including it in the on--shell action gives as a final
result:
\begin{equation}
\delta S_{\mathrm{on-shell}} = \frac{L^2}{2 \kappa_4^2} \int d^3 x L^2 \alpha^3 A(x) \delta \phi_0(x) 
\end{equation}
Therefore, we have our final result:
\begin{equation}
\langle \mathcal{O}_2 (x) \rangle = \frac{1}{d \beta} \frac{\delta S_{\mathrm{on-shell}}}{\delta \phi_0(x)} = \frac{L^2}{2 \kappa_4^2} L^2 \alpha^3 A(x) = \frac{L^2}{2 \kappa_4^2}  \alpha^2 \Psi_2 X(x) \ ,
\end{equation}
where $d$ represents the fact that we want to study the operator in
terms of unit length in the $y$--direction.  If we wish to only
consider the overall scale of the operator ({\it i.e.}, dropping the
$x$ dependence), we can simply write:
\begin{equation}
\langle \mathcal{O}_2 \rangle =  \frac{L^2}{2 \kappa_4^2}  \alpha^2 \Psi_2 \ .
\end{equation}

\providecommand{\href}[2]{#2}\begingroup\raggedright\endgroup

\end{document}